\documentclass[twocolumn]{article}%
\usepackage{amsfonts}
\usepackage{amsmath}
\usepackage{amssymb}
\usepackage{graphicx}%
\providecommand{\U}[1]{\protect\rule{.1in}{.1in}}
\newtheorem{theorem}{Theorem}

\newtheorem{example}[theorem]{Example}

\newtheorem{lemma}[theorem]{Lemma}

\newtheorem{remark}[theorem]{Remark}

\begin{document}

\title{Self-teleportation and its application on LOCC estimation and other tasks}
\author{Keiji Matsumoto\\National Institute of Informatics, 2-1-2, Hitotsubashi, Chiyoda-ku, 101-8430}
\maketitle
\begin{abstract}
A way to characterize quantum nonlocality is to see difference in the figure
of merit between LOCC optimal protocol and globally optimal protocol in doing
certain task, e.g., state estimation, state discrimination, cloning and
broadcasting. Especially, we focus on the case where $n$ tensor of unknown
states.

Our conclusion is that separable pure states are more non-local than entangled
pure states. More specifically, the difference in the figure of the merit is
exponentially small if the state is entangled,
and the exponent is log of the largest Schmidt coefficient. On the other hand,
in many cases, estimation of separable states by LOCC is worse than the global
optimal estimate by $O\left(  \frac{1}{n}\right)  $.

To show that the gap is exponentially small for entangled states, we propose
self-teleportation protocol as the key component of construct of LOCC
protocols. Objective of the protocol is to transfer Alice's part of quantum
information by LOCC, using intrinsic entanglement of
$\left\vert \phi_{\theta}\right\rangle ^{\otimes n}$
without using any extra resources. This protocol
itself is of interest in its own right.
\end{abstract}

\section{Introduction}

A way to characterize quantum nonlocality is to see difference in the figure
of merit between LOCC optimal protocol and globally optimal protocol in doing
certain task, e.g., state estimation, state discrimination, cloning and
broadcasting. Especially, we focus on the case where $n$ tensor of unknown
states: Alice and Bob starts from $\left\vert \phi_{\theta}\right\rangle
^{\otimes n}$ with unknown $\theta$ and $\left\vert \phi_{\theta}\right\rangle
$'s being a member of state family $\left\{  \left\vert \phi_{\theta
}\right\rangle \right\} _{\theta\in\Theta}$. \ 

Our conclusion is that separable pure states are more non-local than entangled
pure states. More specifically, the difference in the figure of the merit is
exponentially small if $\left\vert \phi_{\theta}\right\rangle $ is entangled,
and the exponent is log of the largest Schmidt coefficient. On the other hand,
in many cases, estimation of separable states by LOCC is worse than the global
optimal estimate by $O\left(  \frac{1}{n}\right)  $.

In the past, there had been many works on LOCC state detection. So far as I
know of, there had been no substantial work about LOCC estimation of
continuous unknown parameter. Past results on state detections are either
about very specific case or the statement is rather weak. (see, for example
\cite{OwariHayashi:2006:2}, and references therein.) For example, conditions
for the optimal measurement,  the conditions for perfect detection, upperbound
to the figure of the merit, and so on.  In this work, by introducing
asymptotic point of view, we can cover all the entangled pure states, and
statement about the figure of the merit is had turned out to be the same as
the global optimal measurement, which had been studied in detail.

To show that the gap is exponentially small for entangled states, we propose
self-teleportation protocol as the key component of construct of LOCC
protocols. Objective of the protocol is to transfer Alice's part of quantum
information by LOCC, using intrinsic entanglement of $\left\vert \phi_{\theta
}\right\rangle ^{\otimes n}$ without using any extra resources. This protocol
itself is of interest in its own right.

To show the gap is large for separable pure states, we develop general theory
of LOCC estimation of tensor product states, and show the sufficient
conditions for $O\left(  \frac{1}{n}\right)  $ gap to be observed between LOCC
optimum and the global optimum.

\section{Self-teleportation}

\subsection{objective}

Suppose Alice and Bob share $n$ copies of a unknown $d\times d$ bipartite pure
state $\left\vert \phi\right\rangle _{AB}$. The objective is to transfer the
Alice's quantum information to Bob's local space by LOCC without extra
resources, with exponentially high fidelity:%
\[
\left\vert \phi\right\rangle \left\langle \phi\right\vert ^{\otimes n}%
\in\mathcal{S}\left(  \mathcal{H}_{A}^{\otimes n}\otimes\mathcal{H}%
_{B}^{\otimes n}\right)  \underset{\text{LOCC}}{\longrightarrow}\rho_{\phi
}^{n}\in\mathcal{S}\left(  \mathcal{H}_{B}^{\otimes n}\otimes\mathcal{H}%
_{B^{\prime}}^{\otimes n}\right)  ,
\]
and
\[
\left\langle \phi^{\otimes n}\right\vert \rho_{\phi}^{n}\left\vert
\phi^{\otimes n}\right\rangle \geq1-O\left(  \,\left(  p_{1}^{\phi}\right)
^{n}\,\right)  ,
\]
where $\mathcal{H}_{A}^{\otimes n}$ and $\mathcal{H}_{B}^{\otimes n}%
\otimes\mathcal{H}_{B^{\prime}}^{\otimes n}$ is at Alice and Bob's side,
respectively, and $p_{1}^{\phi}\geq p_{2}^{\phi}\geq\cdots\geq p_{d}^{\phi}$
are the Schmidt coefficient of $\left\vert \phi\right\rangle $.

In case $\left\vert \phi\right\rangle $ is entangled, $p_{1}^{\phi}<1$, and
the fidelity is exponentially close to $1$, while \ it equals 0 for a tensor
product state.

\subsection{A standard form of an ensemble of identical bipartite pure states}

Note $|\phi\rangle^{\otimes n}$ is invariant by the reordering of copies, or
the action of the symmetric group $S_{n}$. Action of the symmetric group
occurs a decomposition of the tensored space $\mathcal{H}^{\otimes n}$
~\cite{Weyl},
\[
\mathcal{H}^{\otimes n}=\bigoplus_{\lambda:\lambda\vdash n}\mathcal{W}%
_{\lambda},\;\mathcal{W}_{\lambda}:=\mathcal{U}_{\lambda}\otimes
\mathcal{V}_{\lambda},
\]
where $\mathcal{U}_{\lambda}$ and $\mathcal{V}_{\lambda}$ is an irreducible
space of the tensor representation of $\mathrm{SU}(d)$, and the representation
of $S_{n}$, respectively, and $\lambda=(\lambda_{1},\ldots,\lambda_{d})$
($\lambda_{i}\geq\lambda_{i+1}\geq0$, $\sum_{i=1}^{d}\lambda_{i}=n$) is called
\textit{Young index}, which $\mathcal{U}_{\lambda}$ and $\mathcal{V}_{\lambda
}$ uniquely corresponds to. We denote by $\mathcal{U}_{\lambda,A}$,
$\mathcal{V}_{\lambda,A}$, and $\mathcal{U}_{\lambda,B}$, $\mathcal{V}%
_{\lambda,B}$ the irreducible component of $\mathcal{H}_{A}^{\otimes n}$ and
$\mathcal{H}_{B}^{\otimes n}$ , respectively. Also, $\mathcal{W}_{\lambda,A}$
and $\mathcal{W}_{\lambda,B}$ are defined analogously.

In terms of \ this decomposition, $\left\vert \phi\right\rangle ^{\otimes n} $
can be written as%

\begin{equation}
\left\vert \phi\right\rangle ^{\otimes n}=\bigoplus_{\lambda:\lambda\vdash
n}a_{\lambda}\left\vert \phi_{\lambda}\right\rangle \left\vert \Phi_{\lambda
}\right\rangle , \label{decomposition}%
\end{equation}
where $\left\vert \phi_{\lambda}\right\rangle \in\mathcal{U}_{\lambda
,A}\otimes\mathcal{U}_{\lambda,B}$, and $\left\vert \Phi_{\lambda
}\right\rangle \in\mathcal{V}_{\lambda,A}\otimes\mathcal{V}_{\lambda,B}$.
While $a_{\lambda}$ and $\left\vert \phi_{\lambda}\right\rangle $ are
dependent on $\left\vert \phi\right\rangle $, $\left\vert \Phi_{\lambda
}\right\rangle $ is a maximally entangled state which does not depend on
$\left\vert \phi\right\rangle $,%
\[
\left\vert \Phi_{\lambda}\right\rangle :=\frac{1}{\sqrt{d_{\lambda}}}%
\sum_{i=1}^{d_{\lambda}}\left\vert f_{A,i}^{\lambda}\right\rangle \left\vert
f_{B,i}^{\lambda}\right\rangle .
\]
Here, $\left\{  \left\vert f_{A,i}^{\lambda}\right\rangle \right\}  $ and
$\left\{  \left\vert f_{B,i}^{\lambda}\right\rangle \right\}  $ is a CONS of
$\mathcal{V}_{\lambda,A}$ and $\mathcal{V}_{\lambda,B}$, respectively, and
$d_{\lambda}:=\dim\mathcal{V}_{\lambda}$.

\subsection{Protocol and performance}

Rough description of our protocol is as follows. Alice teleports her part of
$\left\vert \phi_{\lambda}\right\rangle $ using $\left\vert \Phi_{\lambda
}\right\rangle $. $\left\vert \Phi_{\lambda}\right\rangle $'s will be gone,
but can be reconstructed by Bob since it is independent of $\left\vert
\phi\right\rangle $. Schmidt rank, or equivalently $d_{\lambda}$ is small than
$\dim\mathcal{U}_{\lambda}$ for most of $\lambda$, and our fidelity of success
is exponentially close to 1.

Note that the following protocol does not work: Upon measurement of
$\mathcal{W}_{\lambda,A}\otimes\mathcal{W}_{\lambda,B}$ (In the paper, the
projector will be denoted by the same symbol as its support.), teleport
$\left\vert \phi_{\lambda}\right\rangle $ using $\left\vert \Phi_{\lambda
}\right\rangle $. In such a protocol, coherence between the subspaces
corresponding to different values of $\lambda$ will be destroyed. To keep the
coherence, we use measurement which does not distinguish the Young index
$\lambda$. Consider the following operators.
\[
A_{\left\{  U_{\lambda}\right\}  }:=\bigoplus_{\lambda\in\mathrm{A}_{n}}%
\sqrt{d_{\lambda}}\sum_{i=1}^{\dim\mathcal{U}_{\lambda}}\left\langle
e_{A,i}^{\lambda}\right\vert \,\left\langle f_{A,i}^{\lambda}\right\vert
U_{\lambda}^{\dagger}\,.
\]
Here, $\mathrm{A}_{n}:=\left\{  \lambda:\dim\mathcal{U}_{\lambda}\leq
d_{\lambda}\right\}  $, $\left\{  \left\vert e_{A,i}^{\lambda}\right\rangle
\right\}  $ is a CONS of $\mathcal{U}_{A,\lambda}$, and $U_{\lambda}$ is runs
all over the elements of $\mathrm{U}\left(  \mathcal{V}_{A,\lambda}\right)  $
(NB not $\mathrm{SU}\left(  \mathcal{V}_{A,\lambda}\right)  $). Observe that
$\int U_{\lambda}\mathrm{d}U_{\lambda}=0$ where $\mathrm{d}U_{\lambda}$ is an
invariant measure in $\mathrm{U}(\mathcal{U}_{\lambda})$ with the
normalization $\int\mathrm{d}U_{\lambda}=1$, since $-U_{\lambda}$ is also in
$\mathrm{U}(\mathcal{V}_{\lambda})$. Due to this and the Shur's
lemma\thinspace\ref{lem:shur} and \thinspace\ref{lem:decohere}, we have
\[
\int A_{\left\{  U_{\lambda}\right\}  }^{\dagger}A_{\left\{  U_{\lambda
}\right\}  }\prod_{\lambda\in\mathrm{A}_{n}}\mathrm{d}U_{\lambda}%
=\bigoplus_{\lambda\in\mathrm{A}_{n}}\mathcal{W}_{\lambda,A}\otimes
\mathcal{W}_{\lambda,B}.
\]

\begin{description}
\item[(I)] Alice and Bob project the state onto the subspace $\bigoplus
_{\lambda\in\mathrm{A}_{n}}\mathcal{W}_{\lambda,A}$ and $\bigoplus_{\lambda
\in\mathrm{A}_{n}}\mathcal{W}_{\lambda,B}$ If both of them succeed, they proceed.

\item In fact, from (\ref{decomposition}), if one of them succeeds, both of
them succeeds.

\item The success probability equals $\sum_{\lambda\in\mathrm{A}_{n}%
}a_{\lambda}^{2}$.

\item[(II)] Alice Apply the measurement corresponding to $\left\{  A_{\left\{
U_{\lambda}\right\}  }\right\}  $, and send the measurement outcome $\left\{
U_{\lambda}\right\}  _{\lambda:\lambda\vdash n}$ to Bob. \ After this, Bob has
the state%
\[
\bigoplus_{\lambda\in\mathrm{A}_{n}}a_{\lambda}\sum_{i=1}^{\dim\mathcal{U}%
_{\lambda}}\left\langle e_{A,i}^{\lambda}\right\vert \left.  \phi_{\lambda
}\right\rangle \overline{U_{\lambda}}\left\vert f_{B,i}^{\lambda}\right\rangle
.
\]

\item[(III)] Bob applies the recovery operation $\bigoplus_{\lambda
\in\mathrm{A}_{n}}\mathbf{1}_{\mathcal{U}_{B}}\otimes U_{\lambda}^{T}$, to
obtain $\bigoplus_{\lambda\in\mathrm{A}_{n}}a_{\lambda}\left\vert
\phi_{\lambda}\right\rangle $, where $\left\vert \phi_{\lambda}\right\rangle $
is in $\mathcal{U}_{\lambda,B}\otimes\mathcal{U}_{\lambda,B^{\prime}}$.
Finally, he reconstruct $\left\vert \Phi_{\lambda}\right\rangle $ in
$\mathcal{V}_{\lambda,B}\otimes\mathcal{V}_{\lambda,B^{\prime}}$ to obtain
\begin{equation}
\bigoplus_{\lambda\in\mathrm{A}_{n}}a_{\lambda}\left\vert \phi_{\lambda
}\right\rangle \left\vert \Phi_{\lambda}\right\rangle . \label{final-state}%
\end{equation}
The fidelity between the final state and $\left\vert \phi\right\rangle
^{\otimes n}$ , in average, equals $\sum_{\lambda\in\mathrm{A}_{n}}a_{\lambda
}^{2}$.
\end{description}

Using , we can evaluate $\sum_{\lambda\in\mathrm{A}_{n}}a_{\lambda}^{2}$:%
\begin{align*}
&  \sum_{\lambda\in\mathrm{A}_{n}}a_{\lambda}^{2}\geq\\
&  1-\frac{d\left(  2d-3\right)  !}{\left(  d-2\right)  !\left(  d-1\right)
!}\left(  n+1\right)  ^{\frac{d\left(  d+1\right)  }{2}}\left(  p_{1}^{\phi
}\right)  ^{n}%
\end{align*}

\section{Application to LOCC state estimation}

Suppose Alice and Bob share $n$copies of $\left\vert \phi_{\theta
}\right\rangle $, with unknown continuous parameter $\theta\in\Theta
\subset\mathbb{R}^{D}$. Their task is to estimate $\theta$ as accurately as
possible by LOCC. Our interest is the difference in the figure of merit
between LOCC optimal and the global optimal.

Consider the composition of the globally optimal protocol after the globally
optimal measurement, we observe the figure of the merit differs only by
exponentially small amount, if $\left\vert \phi_{\theta}\right\rangle $ is entangled.

More specifically, there are two kinds of error measures which commonly used.
First one is of the form
\begin{equation}
\mathrm{E\,}dist\left(  \theta,\hat{\theta}_{n}\right)  , \label{E(dist)}%
\end{equation}
where $dist\left(  ,\right)  $ is a smooth distance function, and $\mathrm{E}$
stands for the expectation about the random variable $\hat{\theta}_{n}$. The
second one is
\begin{equation}
\beta_{\epsilon,\theta}^{n}:=\Pr\left\{  \left\Vert \hat{\theta}_{n}%
-\theta\right\Vert >\epsilon\right\}  . \label{error-prob}%
\end{equation}

The first leading term of (\ref{E(dist)}) is, if $dist\left(  ,\right)  $ is
smooth enough, $O\left(  \frac{1}{n}\right)  $. The second leading term is,
looking back classical case, $\Omega\left(  n^{-2}\right)  $, and the third
leading term is $\Omega\left(  n^{-\frac{5}{2}}\right)  $, and so on.
Therefore, our LOCC measurement strategy is as good as the given protocol up
to the higher order terms. \ The error measure (\ref{error-prob}) behaves as
follows:
\[
\beta_{\epsilon,\theta}^{n}\sim e^{-n\,O\left(  \epsilon\right)  }.
\]
Especially, they are interested in the case that $\epsilon$ is small. If
$\epsilon$ is small enough for the exponent of of $\beta_{\epsilon,\theta}%
^{n}$ to be smaller than $\log p_{1}^{\phi}$, our protocol is optimal.

\section{Other applications}

\subsection{State detection}

Suppose $\theta$ takes discrete values, and our aim is to estimate the
parameter $\theta$. Such a problem is called `state detection'. Since
$\left\vert \phi_{\theta}\right\rangle $ and $\left\vert \phi_{\theta^{\prime
}}\right\rangle $ are distant by some constant for all $\theta$ and
$\theta^{\prime}$, the error probability drops exponentially as $n$ increases.
If its exponent is smaller than the one of the RHS of $\log p_{1}%
^{\phi_{\theta}}$, our LOCC protocol will be as good as the given protocol in
the main part of the error. For example, we discuss the sum of all the
possible errors, given the candidates of the true state $\left\vert
\phi_{\theta}\right\rangle $ $(\theta=1,\cdots,M)$. The error exponent cannot
be smaller than the one for the detection problem $\left\vert \phi_{\theta
}\right\rangle $ versus $\left\vert \phi_{\theta^{\prime}}\right\rangle $.
Therefore, $\max_{\theta\neq\theta^{\prime}}\left\vert \left\langle
\phi_{\theta}\right\vert \left.  \phi_{\theta^{\prime}}\right\rangle
\right\vert ^{2}\geq\max_{\theta}p_{1}^{\phi_{\theta}}$ is the sufficient
condition for the optimal LOCC measurement to achieve the global optimum. This
condition holds if $\left\vert \phi_{\theta}\right\rangle $ are distant from
tensor product states.

\subsection{Cloning, broadcast}

Suppose $\theta$ is continuous, and the family $\left\{  \left\vert
\phi_{\theta}\right\rangle \right\}  $ is the totality of the pure states in
d-dimensional Hilbert space. Now, our task is to make $\ m$ copies of
$\left\vert \phi_{\theta}\right\rangle _{BB^{\prime}}$ from $n$ copies of
$\left\vert \phi_{\theta}\right\rangle _{AB}$. The optimal Fidelity of $n$ to
$m$ cloning and broadcast with global operations is $\left(  r\right)
^{d-1}+O\left(  \left(  \frac{1}{n}\right)  \right)  $ and $1+\frac{r-1}%
{rn}+O\left(  \frac{1}{n^{2}}\right)  $ , with $m=rn$ \cite{WernerClone},
\cite{KeylWernerClone}. Seeing these terms, we can observe the first several
terms are obviously larger than exponential order. Hence, our teleportation
scheme does not degrade these terms.

\section{Asymptotic theory estimation of pure separable states}

If $\left\vert \phi_{\theta}\right\rangle =\left\vert \phi_{A,\theta
}\right\rangle \left\vert \phi_{B,\theta}\right\rangle $, we cannot use the
self-teleportation. Hence, this case has to studied separately.  In this
section, we show some sufficient conditions for $O(1/n)$ gap to exist.

\subsection{Asymptotic theory of estimation of pure states}

Given a unknown system, a statistician, assuming that the state corresponds to
a member of a family $\left\{  \rho_{\theta}\,;\theta\in\Theta\subset%
%TCIMACRO{\U{211d} }%
%BeginExpansion
\mathbb{R}
%EndExpansion
^{d}\right\}  $ \ of density matrix, applies a measurement $M$, obtain data,
and calculate the estimate $\hat{\theta}$ of $\theta$ based on the measurement
result. In asymptotic setting, we assume that $\rho_{\theta}^{\otimes n}$ is
given, and the measurement $M^{n}$ may act correctively on $\rho_{\theta
}^{\otimes n}$. A common error measure is (\ref{E(dist)}), where $dist\left(
\cdot,\cdot\right)  $ is smooth enough:
\[
dist\left(  \rho_{\theta},\rho_{\theta+\mathrm{d\,}\theta}\right)  =\sum
_{i,j}G_{ij}\mathrm{d}\theta^{i}\mathrm{d}\theta^{j}+o\left(  \mathrm{d}%
\theta\right)  ^{2}.
\]
With such natural setting, the first leading term of (\ref{E(dist)}) is
$O\left(  1/n\right)  $, and our interest is to minimize the coefficient.

It is known that the optimal coefficient writes
\begin{equation}
\inf_{\left\{  M^{n}\right\}  }\varlimsup_{n\rightarrow\infty}n\mathrm{Tr}%
\,G\left(  J_{\theta}^{M^{n}}\right)  ^{-1}. \label{CR-type}%
\end{equation}
Here $G=\left[  G_{i,j}\right]  $ is a real positive symmetric matrix,
$\mathrm{Tr}$ is trace over $%
%TCIMACRO{\U{211d} }%
%BeginExpansion
\mathbb{R}
%EndExpansion
^{d}$, and $J_{\theta}^{M^{n}}$ is the \textit{Fisher information matrix} of
$p_{\theta}^{M^{n}}\left(  x\right)  :=\mathrm{tr}\,\rho_{\theta}^{\otimes
n}M_{x}$, which is defined by
\begin{equation}
\left[  J_{\theta}^{M^{n}}\right]  _{i,j}:=\sum_{x}p_{\theta}^{M^{n}}\left(
x\right)  \left(  \frac{\partial\log p_{\theta}^{M^{n}}\left(  x\right)
}{\partial\theta^{i}}\,\frac{\partial\log p_{\theta}^{M^{n}}\left(  x\right)
}{\partial\theta^{j}}\,\right)  . \label{def-fisher-meas}%
\end{equation}

As for pure state models, we can explicitly characterize the Fisher
information of the asymptotically optimal measurement \cite{Matsumoto:2002}.
First notable fact is that the collective measurement is not effective:
\[
\min_{M^{n}}n\mathrm{Tr}\,G\left(  J_{\theta}^{M^{n}}\right)  ^{-1}=\min
_{M}\mathrm{Tr}\,G\left(  J_{\theta}^{M}\right)  ^{-1},
\]
where $M$ in the RHS is a measurement acting on the single copy $\left\vert
\phi\right\rangle $. In other words, defining
\[
\mathfrak{J}_{\theta}^{n}:=\left\{  \frac{1}{n}J_{\theta}^{M^{n}}%
\,;M^{n}\,\text{: arbitrary collective meas. }\right\}  .
\]
we have
\[
\mathfrak{J}_{\theta}^{n}=\mathfrak{J}_{\theta}^{1}.
\]

Below, more quantitative results are in order. Define the matrix
\[
J_{\theta,i,j}^{S}:=\Re\left\langle l_{\theta,i}\right.  \left\vert
l_{\theta,j}\right\rangle
\]
and
\[
\tilde{J}_{\theta,i,j}:=\Im\left\langle l_{\theta,i}\right.  \left\vert
l_{\theta,j}\right\rangle ,
\]
where
\[
\left\vert l_{\theta,i}\right\rangle :=\frac{1}{2}\left\{  \frac
{\partial\left\vert \phi_{\theta}\right\rangle }{\partial\theta^{i}%
}-\left\langle \phi_{\theta}\right\vert \frac{\partial}{\partial\theta^{i}%
}\left\vert \phi_{\theta}\right\rangle \left\vert \phi_{\theta}\right\rangle
\right\}  .
\]
These quantities are known to have tight connection with Berry's geometric
phase. Namely, both the line integral of $\left\vert l_{\theta,i}\right\rangle
$ along a closed curve and the area integral of $\tilde{J}_{\theta,i,j}$
\ over the surface enclosed by the curve equals Berry phase.

The eigenvalue of $J_{\theta}^{S-1}\tilde{J}_{\theta}$ is in the form of
$\pm\beta_{\theta,j}$ with \
\[
0\leq\beta_{\theta,j}\leq1.
\]
$\beta_{\theta,j}$ are invariant by the change of the coordinate, and is
closely related to a natural complex structure of the space of pure states.
Indeed, $\arccos\beta_{\theta,j}$ are called multiple Kaehler angles.
$\arccos\beta_{\theta,j}=0$ ($\forall j$) means that the state family is a
complex submanifold.

Below, we mainly treat the case $\dim\theta=2$. In this case, we denote
$\beta_{\theta,1}$ by $\beta_{\theta}$. Suppose that $dist\left(  \cdot
,\cdot\right)  $ is the Bure's distance,
\begin{align*}
dist\left(  \left\vert \phi_{\theta}\right\rangle ,\left\vert \phi
_{\theta+\mathrm{d}\theta}\right\rangle \right)   &  =\sqrt{1-\left\vert
\left\langle \phi_{\theta}\right.  \left\vert \phi_{\theta+\mathrm{d}\theta
}\right\rangle \right\vert ^{2}}\\
&  =\frac{1}{2}\sum_{i,j}J_{\theta,i,j}^{S}\mathrm{d}\theta^{i}\mathrm{d}%
\theta^{j}+o\left(  \mathrm{d}\theta\right)  ^{2}.
\end{align*}
If we use this measure of the error,
\begin{align}
&  \inf_{\left\{  M^{n}\right\}  }\varlimsup_{n\rightarrow\infty}%
n\sqrt{1-\left\vert \left\langle \phi_{\theta}\right.  \left\vert \phi
_{\hat{\theta}^{n}}\right\rangle \right\vert ^{2}}\nonumber\\
&  =\min_{M}\mathrm{Tr}\,J_{\theta}^{S}\left(  J_{\theta}^{M}\right)
^{-1}\nonumber\\
&  =\sum_{j}\frac{4}{1+\sqrt{1-\beta_{\theta,j}^{2}}}. \label{CR-SLD-Weight}%
\end{align}
Hence, the error is monotone increasing in $\beta_{\theta,j}$.

\subsection{The asymptotic estimation of tensor product pure state by LOCC}

Suppose
\[
\rho_{\theta}=\rho_{A,\theta}\otimes\rho_{B,\theta},
\]
with $\rho_{A,\theta}$ and $\rho_{B,\theta}$ is supported on $\mathcal{H}_{A}
$ and $\mathcal{H}_{B}$, respectively, and define $J_{\theta}^{M^{A,n}}$ by
replacing $M^{n}$ and $\rho_{\theta}^{\otimes n}$ in (\ref{def-fisher-meas})
by $M^{A,n}$ and $\rho_{A,\theta}^{\otimes n}$, respectively. $J_{\theta
}^{M^{B,n}}$ is also similarly defined. As in appendix, slight modification of
the proof of lemma\thinspace1 in \cite{HayashiMatsumoto:2002} leads to:

\begin{lemma}
\label{lem:loccM=tensorM} If $M^{n}$ is a LOCC measurement, there is a
measurement $M^{A,n}$ and $M^{B,n}$ acting on $\mathcal{H}_{A}^{\otimes n}$
and $\mathcal{H}_{B}^{\otimes n}$ , respectively, and satisfies
\[
J_{\theta}^{M^{n}}=J_{\theta}^{M^{A,n}}+J_{\theta}^{M^{B,n}}.
\]
Here $J_{\theta}^{M^{A,n}}$ is defined by replacing $M^{n}$ and $\rho_{\theta
}^{\otimes n}$ in (\ref{def-fisher-meas}) by $M^{A,n}$ and $\rho_{A,\theta
}^{\otimes n}$, respectively. $J_{\theta}^{M^{B,n}}$ is also similarly defined.
\end{lemma}

The lemma will be proved in subsection\thinspace\ref{subsec:pf-loccM=tensorM}.
This lemma means that LOCC measurement $M^{n}$ can be replaced by the local
measurements $M^{A,n}$ and $M^{B,n}$, so far as the Fisher information is concerned.

Define%

\begin{align*}
&  \mathfrak{J}_{\theta}^{LOCC,n}\\
&  :=\left\{  \frac{1}{n}J_{\theta}^{M^{n}};%
\begin{array}
[c]{c}%
M^{n}\text{:LOCC in A-B split, }\\
\text{collective over }n\text{ copies}\,
\end{array}
\right\}  ,
\end{align*}
this lemma can be expressed as%
\begin{align*}
&  \mathfrak{J}_{\theta}^{LOCC,n}\\
&  =\left\{  J\,;\,J=J^{A}+J^{B},J^{A}\in\mathfrak{J}_{\theta}^{A,n}%
\,,J^{B}\in\mathfrak{J}_{\theta}^{B,n}\,\right\}  .
\end{align*}

In pure state family case, as we seen above, $\mathfrak{J}_{\theta}%
^{n}=\mathfrak{J}_{\theta}^{1}$. Hence, combined with lemma\thinspace
\ref{lem:loccM=tensorM}, we can conclude that $\mathfrak{J}_{\theta}%
^{LOCC,n}=\mathfrak{J}_{\theta}^{LOCC,1}$ , meaning that being collective over
the copies is not useful also LOCC case, either.

Define $\left\vert l_{\theta,i}^{A}\right\rangle $, $J_{\theta}^{S,A}$,
$\tilde{J}_{\theta}^{A}$, $\beta_{\theta,j}^{A}$ and $\left\vert l_{\theta
,i}^{B}\right\rangle $, $J_{\theta}^{S,B}$, $\tilde{J}_{\theta}^{B}$,
$\beta_{\theta,j}^{B}$ by replacing $\left\vert \phi_{\theta}\right\rangle $
by $\left\vert \phi_{\theta}^{A}\right\rangle $ and $\left\vert \phi_{\theta
}^{B}\right\rangle $, respectively. Observe that%
\begin{align}
\left\vert l_{\theta,i}\right\rangle  &  =\left\vert l_{\theta,i}%
^{A}\right\rangle \left\vert \phi_{\theta}^{B}\right\rangle +\left\vert
\phi_{\theta}^{A}\right\rangle \left\vert l_{\theta,i}^{B}\right\rangle
,\nonumber\\
J_{\theta}^{S}  &  =J_{\theta}^{S,A}+J_{\theta}^{S,B},\label{Js:add}\\
\tilde{J}_{\theta}  &  =\tilde{J}_{\theta}^{,A}+\tilde{J}_{\theta}^{B}.
\label{tildeJ:add}%
\end{align}
Further, we suppose $\dim\theta=2$ and
\[
J_{\theta}^{S,A}=aA,\text{ }J_{\theta}^{S,B}=bA
\]
with $A=A^{T}\geq0$. This means with proper coordinate system, $J_{\theta
}^{S,A}=a\mathbf{1}$ and $J_{\theta}^{S,B}=b\mathbf{1}$. Hence, we have
\begin{equation}
\beta_{\theta}=\frac{a\beta_{\theta}^{A}\pm b\beta_{\theta}^{B}}{a+b},
\label{beta-add}%
\end{equation}
and
\begin{align*}
&  \max_{J\in\mathfrak{J}_{\theta}^{1}}\mathrm{Tr}\,J_{\theta}^{S-1}J\\
&  =1+\sqrt{1-\left(  \frac{a\beta_{\theta}^{A}\pm b\beta_{\theta}^{B}}%
{a+b}\right)  ^{2}}\\
&  \geq\frac{1}{a+b}\left(
\begin{array}
[c]{c}%
a\left(  1+\sqrt{1-\left(  \beta_{\theta}^{A}\right)  ^{2}}\right) \\
+b\left(  1+\sqrt{1-\left(  \beta_{\theta}^{B}\right)  ^{2}}\right)
\end{array}
\right)  ,
\end{align*}
where the identity holds if and only if $\beta_{\theta}^{A}=\beta_{\theta}%
^{B}$ and the $+$-sign in (\ref{beta-add}) is the case. On the other hand,
\begin{align*}
&  \max_{J\in\mathfrak{J}_{\theta}^{LOCC,1}}\mathrm{Tr}\,J_{\theta}^{S-1}J\\
&  =\max_{J\in\mathfrak{J}_{\theta}^{A,1}}\mathrm{Tr}\,J_{\theta}^{S-1}%
J+\max_{J\in\mathfrak{J}_{\theta}^{B,1}}\mathrm{Tr}\,J_{\theta}^{S-1}J\\
&  =\frac{a}{a+b}\max_{J\in\mathfrak{J}_{\theta}^{A,1}}\mathrm{Tr}\,\left(
J_{\theta}^{S,A}\right)  ^{-1}J+\frac{b}{a+b}\max_{J\in\mathfrak{J}_{\theta
}^{B,1}}\mathrm{Tr}\,\left(  J_{\theta}^{S,B}\right)  ^{-1}J\\
&  =\frac{1}{a+b}\left\{
\begin{array}
[c]{c}%
a\left(  1+\sqrt{1-\left(  \beta_{\theta}^{A}\right)  ^{2}}\right) \\
+b\left(  1+\sqrt{1-\left(  \beta_{\theta}^{B}\right)  ^{2}}\right)
\end{array}
\right\}  .
\end{align*}
Therefore, we have%

\[
\max_{J\in\mathfrak{J}_{\theta}^{1}}\mathrm{Tr}\,J_{\theta}^{S-1}J\geq
\max_{J\in\mathfrak{J}_{\theta}^{LOCC,1}}\mathrm{Tr}\,J_{\theta}^{S-1}J.
\]
Due to \cite{Matsumoto:2002}, the maximum of $\mathrm{Tr}\,J_{\theta}^{S-1}J$
\ and the minimum of $\mathrm{Tr}\,J_{\theta}^{S}J^{-1}$is achieved by the
same matrix, a constant multiple of $J_{\theta}^{S}$. Therefore, we have
\[
\min_{J\in\mathfrak{J}_{\theta}^{1}}\mathrm{Tr}\,J_{\theta}^{S}J^{-1}%
>\max_{J\in\mathfrak{J}_{\theta}^{LOCC,1}}\mathrm{Tr}\,J_{\theta}^{S-1}J
\]
unless $\beta_{\theta}^{A}=\beta_{\theta}^{B}$ and the $+$-sign in
(\ref{beta-add}) is the case. Given $J_{\theta}^{S,A}$ and $J_{\theta}^{S,B}$,
the gap between the both ends becomes largest when $\beta_{\theta}^{A}%
=\beta_{\theta}^{B}=1$ and $\beta_{\theta}=0$.

\begin{example}
Let $\left\vert \phi_{\theta}^{A}\right\rangle $ qubit states, and $\left\vert
\phi_{\theta}^{B}\right\rangle $ be its complex complement, i.e.,%
\[
\left\vert \phi_{\theta}^{A}\right\rangle =\left[
\begin{array}
[c]{c}%
e^{-\sqrt{-1}\frac{\theta^{2}}{2}}\cos\frac{\theta^{1}}{2}\\
e^{\sqrt{-1}\frac{\theta^{2}}{2}}\sin\frac{\theta^{1}}{2}%
\end{array}
\right]  ,\,\left\vert \phi_{\theta}^{B}\right\rangle =\left[
\begin{array}
[c]{c}%
e^{\sqrt{-1}\frac{\theta^{2}}{2}}\cos\frac{\theta^{1}}{2}\\
e^{-\sqrt{-1}\frac{\theta^{2}}{2}}\sin\frac{\theta^{1}}{2}%
\end{array}
\right]  ,
\]
It is easy to check
\[
J_{\theta}^{S,A}=J_{\theta}^{S,B},
\]
and
\begin{align*}
\beta_{\theta}^{A}  &  =\beta_{\theta}^{B}=1,\\
\beta_{\theta}  &  =0.
\end{align*}
This example is so called 'unti-copy'. With LOCC, one can never make use of
the effect of having unit-copy.
\end{example}

A natural generalization of unit-copy would be \bigskip the state family with
$\beta_{\theta,j}=0$ ($\forall j$). This is necessary and sufficient
\cite{Matsumoto:2002} for being
\[
\mathfrak{J}_{\theta}^{1}=\left\{  J\,\,;\,J\leq J_{\theta}^{S}\right\}  .
\]

Due to (\ref{Js:add}) and lemma\thinspace\ref{lem:loccM=tensorM},
$\mathfrak{J}_{\theta}^{1}=\mathfrak{J}_{\theta}^{LOCC,1}$ occurs if and only
if $\mathfrak{J}_{\theta}^{A,1}=\left\{  J\,\,;\,J\leq J_{\theta}%
^{S,A}\right\}  $ and $\mathfrak{J}_{\theta}^{B,1}=\left\{  J\,\,;\,J\leq
J_{\theta}^{S,B}\right\}  $. $\ $This condition is equivalent to
$\beta_{\theta,j}^{A}=\beta_{\theta,j}^{B}=0$ ($\forall j$). Hence, if
$\beta_{\theta,j}=0$ ($\forall j$) and $\beta_{\theta,j}^{A}\neq0$ ($\exists
j$), the gap between LOCC an the globally optimal measurement can be observed.

\begin{remark}
A tricky case is that $\left\vert \phi_{\theta}\right\rangle $ is separable
only for $\theta\in\Theta_{s}\subset\Theta$. Suppose $\Theta$ is an open set
in $%
%TCIMACRO{\U{211d} }%
%BeginExpansion
\mathbb{R}
%EndExpansion
^{d^{\prime}}$, and that that the map $\theta\rightarrow$ $\left\vert
\phi_{\theta}\right\rangle $ is smooth. Then, $\dim\Theta_{s}$ has to be
strictly smaller than $\dim\Theta$. Therefore, if we consider an arbitrary
prior distribution of $\theta$ which can be written as $q\left(
\theta\right)  \mathrm{d}\theta$, with $\mathrm{d}\theta$ being the Lebesgue
measure, it won't contribute to the average of the figure of merit with
respect to the prior distribution.
\end{remark}

\subsection{Proof of lemma\thinspace\ref{lem:loccM=tensorM}}

\label{subsec:pf-loccM=tensorM} The proof is similar to the proof of
lemma\thinspace1 in \cite{HayashiMatsumoto:2002}. Let $x_{t}$ and $y_{t}$ be
the data obtained at $t$th round, and define $x^{t-1}:=x_{1}\cdots x_{t-1}$,
and $y^{t-1}:=y_{1}\cdots y_{t-1}$ . The measurement by Alice and Bob at $t$th
round is denoted by $A_{t}^{x^{t-1}y^{t-1}}$ and $B_{t}^{x^{t-1}y^{t-1}}$, respectively.

The key point is that in optimization (\ref{CR-type}), the measurement $M^{n}$
can depend on the true value of $\theta$. Note that this is not the case for
the estimation scheme minimizing (\ref{E(dist)}). However, to compute the
first order asymptotic term of (\ref{E(dist)}), we only have to solve the
optimization problem (\ref{CR-type}), where the measurement may depend on
unknown parameter $\theta$. The fact that $\theta$ is unknown is reflected in
the fact the Fisher information is a function of the derivative of the
probability distribution with respect to $\theta$.

Suppose also Alice has $\rho_{A,\theta}^{\otimes n}\otimes\rho_{B,\theta_{0}%
}^{\otimes n}$ and \ Bob has $\rho_{A,\theta_{0}}^{\otimes n}\otimes
\rho_{B,\theta}^{\otimes n}$, locally. Instead of doing communication, Alice
applies $A_{t}^{x^{t-1}y^{t-1}}$ to $\rho_{A,\theta}^{\otimes n}$, and
$B_{t}^{x^{t-1}y^{t-1}}$ to $\rho_{B,\theta_{0}}^{\otimes n}$, and Bob also
does the same. If $\theta=\theta_{0}$, Fisher information matrix of this local
measurement scheme is the scheme equals the one of the LOCC measurement scheme
specified by $A_{t}^{x^{t-1}y^{t-1}}$ and $B_{t}^{x^{t-1}y^{t-1}}$, as is
shown below. \ Since the construction of measurement can depend on the value
of $\theta$, we have the lemma.

The measurement scheme corresponding to this optimum solution of
(\ref{CR-type}) is as follows. Alice and Bob measures $\sqrt{n}$ copies of
$\rho_{\theta}$ locally, and exchange the measurement data. They compute
auxiliary estimate $\tilde{\theta}_{n}$ . Believing that this value is true,
Alice and Bob fabricate $\rho_{B,\tilde{\theta}_{n}}^{\otimes n-\sqrt{n}}$ and
$\rho_{A,\tilde{\theta}_{n}}^{\otimes n-\sqrt{n}}$, and applies the LOCC
measurement optimal at $\theta=\tilde{\theta}_{n}$ to $\left(  \rho_{A,\theta
}\otimes\rho_{B,\tilde{\theta}_{n}}\right)  ^{\otimes n-\sqrt{n}}$ and
$\left(  \rho_{A,\tilde{\theta}_{n}}\otimes\rho_{B,\theta}\right)  ^{\otimes
n-\sqrt{n}}$, respectively. Finally, they exchange the measurement data, and
compute the estimate.

Below, we assume $\dim\theta=1$ for simplicity, but general case is a trivial
generalization. If the LOCC measurement $M^{n}$ is realized by LOCC with $t$
rounds of exchange of classical communication, denoting
\begin{align*}
p_{\theta}\left(  x^{t}|y^{t-1}\right)   &  =\Pr_{\theta}\left\{  x^{t}%
|y^{t}\right\} \\
q_{\theta}\left(  y^{t}|x^{t-1}\right)   &  =\Pr_{\theta}\left\{  y^{t}%
|x^{t}\right\}  ,
\end{align*}
we have%
\begin{align*}
&  \left.  J_{\theta}^{M^{n}}\right\vert _{\theta=\theta_{0}}\\
&  =\sum_{x^{t},y^{t}}\left[
\begin{array}
[c]{c}%
p_{\theta}\left(  x^{t}|y^{t-1}\right)  q_{\theta}\left(  y^{t}|x^{t-1}\right)
\\
\times\left(  \frac{\mathrm{d}}{\mathrm{d}\theta}\ln p_{\theta}\left(
x^{t}|y^{t-1}\right)  q_{\theta}\left(  y^{t}|x^{t-1}\right)  \right)  ^{2}%
\end{array}
\right]  _{\theta=\theta_{0}}\\
&  =\sum_{x^{t},y^{t}}\left[
\begin{array}
[c]{c}%
q_{\theta}\left(  y^{t}|x^{t-1}\right)  \times\\
p_{\theta}\left(  x^{t}|y^{t-1}\right)  \left(  \frac{\mathrm{d}}%
{\mathrm{d}\theta}\ln p_{\theta}\left(  x^{t}|y^{t-1}\right)  \right)  ^{2}%
\end{array}
\right]  _{\theta=\theta_{0}}\\
&  +\sum_{x^{t},y^{t}}\left[
\begin{array}
[c]{c}%
p_{\theta}\left(  x^{t}|y^{t-1}\right)  \times\\
q_{\theta}\left(  y^{t}|x^{t-1}\right)  \left(  \frac{\mathrm{d}}%
{\mathrm{d}\theta}\ln q_{\theta}\left(  y^{t}|x^{t-1}\right)  \right)  ^{2}%
\end{array}
\right]  _{\theta=\theta_{0}}\\
&  +\sum_{x^{t},y^{t}}\left[
\begin{array}
[c]{c}%
p_{\theta}\left(  x^{t}|y^{t-1}\right)  q_{\theta}\left(  y^{t}|x^{t-1}\right)
\\
\times\frac{\mathrm{d}}{\mathrm{d}\theta}\ln q_{\theta}\left(  y^{t}%
|x^{t-1}\right)  \frac{\mathrm{d}}{\mathrm{d}\theta}\ln p_{\theta}\left(
x^{t}|y^{t-1}\right)
\end{array}
\right]  _{\theta=\theta_{0}}\,\,.
\end{align*}

Here, observe the last term equals 0, due to the following reason. We have
\begin{align*}
&  \sum_{x^{t},y^{t}}p_{\theta}\left(  x^{t}|y^{t-1}\right)  q_{\theta}\left(
y^{t}|x^{t-1}\right)  \\
&  \quad\times\frac{\mathrm{d}}{\mathrm{d}\theta}\ln p_{\theta}\left(
x^{t}|y^{t-1}\right)  \frac{\mathrm{d}}{\mathrm{d}\theta}\ln q_{\theta}\left(
y^{t}|x^{t-1}\right)  \\
&  =\sum_{x^{t},y^{t}}\left[
\begin{array}
[c]{c}%
p_{\theta}\left(  x_{t}|x^{t-1}y^{t-1}\right)  p_{\theta}\left(
x^{t-1}|y^{t-2}\right)  \\
\times q_{\theta}\left(  y_{t}|y^{t-1}x^{t-1}\right)  q_{\theta}\left(
y^{t-1}|x^{t-2}\right)  \\
\times\left(
\begin{array}
[c]{c}%
\frac{\mathrm{d}}{\mathrm{d}\theta}\ln p_{\theta}\left(  x_{t}|x^{t-1}%
y^{t-1}\right)  \\
+\frac{\mathrm{d}}{\mathrm{d}\theta}\ln p_{\theta}\left(  x^{t-1}%
|y^{t-2}\right)
\end{array}
\right)  \\
\times\left(
\begin{array}
[c]{c}%
\frac{\mathrm{d}}{\mathrm{d}\theta}\ln q_{\theta}\left(  y_{t}|y^{t-1}%
x^{t-1}\right)  \\
+\frac{\mathrm{d}}{\mathrm{d}\theta}\ln q_{\theta}\left(  y^{t-1}%
|x^{t-2}\right)
\end{array}
\right)
\end{array}
\right]  \\
&  =\sum_{x^{t-1},y^{t-1}}\left[
\begin{array}
[c]{c}%
p_{\theta}\left(  x^{t-1}|y^{t-2}\right)  q_{\theta}\left(  y^{t-1}%
|x^{t-2}\right)  \\
\times\left(  \frac{\mathrm{d}}{\mathrm{d}\theta}\ln p_{\theta}\left(
x^{t-1}|y^{t-2}\right)  \right)  \\
\times\left(  \frac{\mathrm{d}}{\mathrm{d}\theta}\ln q_{\theta}\left(
y^{t-1}|x^{t-2}\right)  \right)
\end{array}
\right]  \\
&  =\sum_{x_{1}y_{1}}p_{\theta}\left(  x_{1}\right)  q_{\theta}\left(
y_{1}\right)  \frac{\mathrm{d}}{\mathrm{d}\theta}\ln p_{\theta}\left(
x_{1}\right)  \frac{\mathrm{d}}{\mathrm{d}\theta}\ln q_{\theta}\left(
y_{1}\right)  \\
&  =0.
\end{align*}
Hence, we have%
\begin{align*}
&  \left.  J_{\theta}^{M^{n}}\right\vert _{\theta=\theta_{0}}\\
&  =\sum_{x^{t},y^{t}}\left[
\begin{array}
[c]{c}%
q_{\theta_{0}}\left(  y^{t}|x^{t-1}\right)  \times\\
p_{\theta}\left(  x^{t}|y^{t-1}\right)  \left(  \frac{\mathrm{d}}%
{\mathrm{d}\theta}\ln p_{\theta}\left(  x^{t}|y^{t-1}\right)  \right)  ^{2}%
\end{array}
\right]  _{\theta=\theta_{0}}\\
&  +\sum_{x^{t},y^{t}}\left[
\begin{array}
[c]{c}%
p_{\theta_{0}}\left(  x^{t}|y^{t-1}\right)  \times\\
q_{\theta}\left(  y^{t}|x^{t-1}\right)  \left(  \frac{\mathrm{d}}%
{\mathrm{d}\theta}\ln q_{\theta}\left(  y^{t}|x^{t-1}\right)  \right)  ^{2}%
\end{array}
\right]  _{\theta=\theta_{0}}%
\end{align*}
Therefore, the first term is the average of the Fisher information of the
probability distribution family $\left\{  \,p_{\theta}\left(  x^{t}%
|y^{t-1}\right)  \,\right\}  _{\theta\in\Theta}$ \ with $y^{t}$ obeying
$q_{\theta_{0}}\left(  y^{t}|x^{t-1}\right)  $. The second term is the similar.

\section{Acknowledgment}

The author is thankful to M. Hayashi for pointing out that the proof of
lemma\thinspace\ref{lem:loccM=tensorM} can be done in the analogous way as the
proof of lemma\thinspace1 in \cite{HayashiMatsumoto:2002}.

\bigskip

\appendix

\section{Group representation theory}

\label{appendixA}

\begin{lemma}
\label{lem:decohere} Let $U_{g}$ and $U_{g}^{\prime}$ be an irreducible
representation of $G$ on the finite-dimensional space $\mathcal{H}$ and
$\mathcal{H}^{\prime}$, respectively. We further assume that $U_{g}$ and
$U_{g}^{\prime}$ are not equivalent. If a linear operator $A$ in
$\mathcal{H}\oplus\mathcal{H}^{\prime}$ is invariant by the transform
$A\rightarrow U_{g}\oplus U_{g}^{\prime}AU_{g}^{\ast}\oplus U_{g}^{^{\prime
}\ast}$ for any $g$, $\mathcal{H}A\mathcal{H^{\prime}}=0$ ~\cite{GW}.
\end{lemma}

\begin{lemma}
\label{lem:shur} (Shur's lemma~\cite{GW}) Let $U_{g}$ be as defined in
lemma~\ref{lem:decohere}. If a linear map $A$ in $\mathcal{H}$ is invariant by
the transform $A\rightarrow U_{g}AU_{g}^{\ast}$ for any $g$, $A=c\mathrm{Id}%
_{\mathcal{H}}$.
\end{lemma}

\section{Representation of symmetric group and SU}

Due to \cite{GW}, we have%

\begin{align}
\dim\mathcal{U}_{\lambda}  &  =\frac{\prod_{i<j}\left(  l_{i}-l_{j}\right)
}{\prod_{i=1}^{d-1}\left(  d-i\right)  !},\label{dim-representation-1}\\
d_{\lambda}  &  =\dim\mathcal{V}_{\lambda}=\frac{n!}{\prod_{i=1}^{d}\left(
\lambda_{i}+d-i\right)  !}\prod_{i<j}\left(  l_{i}-l_{j}\right)  ,
\label{dim-representation-2}%
\end{align}
with $l_{i}:=\lambda_{i}+d-i$. \ It is easy to show%
\begin{equation}
\log\dim\mathcal{U}_{\lambda}\leq d^{2}\log n. \label{dim-zero-rate}%
\end{equation}

Let $a_{\lambda}^{\phi}=\mathrm{Tr}\left\{  \mathcal{W}_{\lambda,A}\left(
\mathrm{Tr}_{B}|\phi\rangle\langle\phi|\right)  ^{\otimes n}\right\}  $ and
the formulas in the appendix of \cite{Ha} says%

\begin{align}
\left\vert \frac{\log d_{\lambda}}{n}-\mathrm{H}\left(  \frac{\lambda}%
{n}\right)  \right\vert  &  \leq\frac{d^{2}+2d}{2n}\log
(n+d),\label{grep-type-1}\\
\sum_{\frac{\lambda}{n}\in\mathrm{R}}a_{\lambda}^{\phi}  &  \leq\left(
n+1\right)  ^{d\left(  d+1\right)  /2}\exp\left\{  -n\min_{\boldsymbol{q}%
\,\in\mathrm{R}}\mathrm{D}\left(  \boldsymbol{q}||\boldsymbol{p}\right)
\right\}  , \label{grep-type-2}%
\end{align}
where $\mathrm{R}$ is an arbitrary closed subset.

\section{}
\end{document}